\documentclass[final,1p,times,sort&compress,12pt]{elsarticle}

\usepackage{amssymb}
\usepackage{amsmath}
\usepackage{siunitx}

\usepackage{amsmath,amssymb}
\usepackage{graphicx}
\usepackage{subcaption}
\usepackage{hyperref}
\usepackage{siunitx}
\usepackage{caption}
\journal{Acta Materialia}

\makeatletter
\def\ps@pprintTitle{%
  \let\@oddhead\@empty
  \let\@evenhead\@empty
  \def\@oddfoot{\reset@font\hfil\thepage\hfil}
  \let\@evenfoot\@oddfoot
}
\makeatother

\begin{document}

\begin{frontmatter}

\title{On stress-assisted boundary migration during recrystallization}

\author[1]{Yubin Zhang\corref{cor1}}
\author[2]{Qiwei Shi}
\author[3]{Guilin Wu}

\cortext[cor1]{Corresponding author: \texttt{yubz@dtu.dk}; Postal address: Produktionstorvet Building 425, 2800 Kgs. Lyngby, Denmark}

\address[1]{Department of Civil and Mechanical Engineering, Technical University of Denmark, DK-2800, Kgs.\ Lyngby, Denmark}
\address[2]{SJTU-Paris Elite Institute of Technology, Shanghai Jiao Tong University, Shanghai 200240, China}
\address[3]{International Joint Laboratory for Light Alloys (MOE), College of Materials Science and Engineering, Chongqing University, Chongqing 400044, China}

\begin{abstract}

This study investigates the boundary migration mechanisms near the sample surface of recrystallizing grains in high-purity Al subjected to cryogenic rolling. Local strain and stress tensors were characterized during \textit{in situ} annealing by combining high-resolution electron backscatter diffraction with microstructure-based digital image correlation strain analysis. The results reveal local residual strains on the order of $10^{-3}$ within the recrystallizing grain, with values several times higher in the adjacent deformed matrix. The residual stresses in recrystallizing grains are a passive response to those developed within the surrounding deformed grains; the latter being strongly influenced by the local geometry and characteristics of dislocation boundaries, as well as by constraints imposed by neighboring grains. No evidence of shear-coupled motion was observed during the recrystallization boundary migration, despite the presence of shear stress across the boundary. In contrast, detailed analysis of the principal strain components reveals a clear correlation between residual strain patterns and boundary migration directions. These findings indicate that recrystallization boundary migration is modulated by the anisotropy of the local internal stress state.

\end{abstract}


\end{frontmatter}

\section{Introduction}

Recrystallizing grains/nuclei developed during annealing of deformed metals and alloys have traditionally been considered as stress-free~\cite{ref1}. However, recent synchrotron micro-diffraction experiments have provided clear evidence of the presence of residual stresses within the recrystallizing grains in partially recrystallized samples~\cite{ref2,ref3,ref4,ref5,ref6}. While a pronounced macroscopic strain pattern, exhibiting a strong correlation between the strain tensor directions and sample directions, has been observed, significant local stress variations exist both within and across individual recrystallizing grains~\cite{ref5}. The maximum stresses can approach the yield stress of the deformed matrix~\cite{ref5}. It has been argued that the development of such strain/stress is attributed to differences in defect density between recrystallizing grains and the surrounding deformed matrix, as well as to the redistribution of medium- to long-range residual stresses within the deformed matrix~\cite{ref7,ref8,ref9}. However, how the residual stresses within recrystallizing grains develop and how they affect local boundary migration during recrystallization remain unclear, primarily due to the limited understanding of the local residual stress tensors within the deformed matrix.

Characterizing residual stress tensors within deformed microstructures is inherently challenging, particularly at high plastic strains where the defect density is substantial. Synchrotron-based techniques, such as three-dimensional X-ray micro-beam Laue diffraction (3D$\mu$XRD, also known as differential aperture X-ray microscopy)~\cite{ref10,ref11} and scanning three-dimensional X-ray diffraction (s-3DXRD)~\cite{ref12,ref13,ref14,ref15,ref16} have been employed to probe the local elastic strain tensors and lattice rotation fields in three dimensions in grains with low defect density~\cite{ref5,ref13,ref14,ref17}. However, their application to heavily deformed microstructures remains limited due to the low signal-to-noise ratio caused by severe lattice distortion.

Alternatively, electron microscopy techniques, such as high-angular-resolution electron backscatter diffraction (HR-EBSD), can be used to characterize local stress tensors through cross-correlation analysis of electron backscatter patterns (EBSPs) (also known as Kikuchi patterns)~\cite{ref18,ref19,ref20,ref21,ref22,ref23}. Most commonly, HR-EBSD determines crystallographic orientation and elastic strain relative to a reference pattern using local subregion-based cross-correlation~\cite{ref19,ref20,ref21}. Global whole-frame-based digital image correlation (DIC) has also been applied for HR-EBSD, providing fast and precise analysis for both simulated and experimental EBSPs~\cite{ref23,ref24,ref25}. More recently, advances have enabled the measurement of absolute residual stresses using global-based DIC with simulated reference patterns by minimizing a cost function formulated from the gradients of the EBSPs~\cite{ref26,ref27}. Although HR-EBSD provides only surface information and may therefore not be fully representative of the bulk, its high spatial resolution offers valuable insights into the local stress and strain distributions within deformed matrix, making it the best available option.

At an intermediate scale, DIC analysis of surface speckle patterns or tracking of surface particles/features allows quantification of surface strain evolution during deformation~\cite{ref28,ref29,ref30,ref31,ref32,ref33}. While this type of surface-feature DIC does not yield the full strain tensor, it provides in-plane strain and displacement fields, complementing the fine-scale information obtained from HR-EBSD. The combined use of surface-feature DIC and HR-EBSD thus offers a powerful framework for linking surface strain evolution with local residual stress states within the microstructure, enabling the investigation of open questions regarding the role of local stresses in governing grain boundary migration behavior during recrystallization.

The latter is particularly important, as there has recently been growing recognition of stress-induced boundary migration mechanisms, such as shear-coupled motion~\cite{ref34,ref35,ref36,ref37,ref38,ref39,ref40,ref41,ref42}. Shear coupling refers to boundary motion driven by shear stress acting across the boundary plane, which in turn causes a relative tangential displacement between the two adjoining grains. Recent simulations have demonstrated that shear coupling can account for the lack of correlation between boundary curvature and migration velocity during grain growth~\cite{ref43,ref44,ref45}. However, it has also been shown experimentally that, in certain cases, boundary motion under an applied stress does not produce measurable shear displacement~\cite{ref46}. Given that local residual stresses have been detected within recrystallizing grains~\cite{ref2,ref3,ref4,ref5,ref6} and may exert shear stresses across these boundaries, it is important to clarify whether these stresses contribute to a shear-coupled motion and influence the local recrystallization boundary migration.

The aim of this study is to explore the mechanisms underlying the formation of local residual stresses and their impact on local boundary migration by analyzing the local migration behavior of recrystallizing boundaries in high-purity (99.996\%) aluminum, tracked during \textit{in situ} annealing conducted inside a scanning electron microscope (SEM). Two recrystallization boundaries (GBs) are examined. While the local boundary migration velocities of the first GB were quantified in an earlier study~\cite{ref47}, the present work introduces a new analysis of local in-plane strains using a microstructure-based DIC approach and correlates these with the local boundary migration patterns. To understand the influence of local residual stresses on local boundary migration, new in-situ experiments were carried out for a second GB, where residual stresses are characterized using a global-DIC-based HR-EBSD method. The results obtained for the two GBs are compared and used to evaluate potential shear-coupled migration mechanisms and the origins of local residual stress development. The implications of local residual stresses and stress-assisted migration mechanisms for the recrystallization process are discussed.

\section{Experimental}

The sample used was rolled pure aluminum with an initial grain size of several millimeters. The rolling was first performed at room temperature to a 50\% reduction in thickness, followed by rolling to a total thickness reduction of 86\% at liquid-nitrogen temperature, in order to suppress dynamic recrystallization during deformation. After rolling, the material was stored in a freezer at $-18\,^{\circ}\mathrm{C}$. Specimens of approximately $2 \times 2 \times 6~\mathrm{mm}^3$ (with rolling direction, RD, along the long side) were prepared and mechanically polished on the longitudinal section, defined by sample normal direction (ND) and RD, following standard metallurgical procedures, and subsequently electropolished using an A2 solution at $0\,^{\circ}\mathrm{C}$ for 45~s under an applied voltage of 13~V.

Two specimens were examined, each used to track specific objectives of local boundary migration during in situ annealing. In-situ annealing was conducted using a heating stage (DEBEN UK Ltd.) mounted inside a Zeiss Supra 35 thermal field-emission-gun scanning electron microscope (SEM). For both in-situ measurements, the specimen was cooled from room temperature to $0\,^{\circ}\mathrm{C}$ inside the SEM to stabilize the microstructure before initial EBSD characterization. The EBSD data were acquired at an accelerating voltage of 15~kV and a sample tilt angle of $70^{\circ}$, using an Oxford Instruments system equipped with HKL Channel 5 control software and a NordlysNano high-resolution CCD detector ($1344 \times 1024$ pixels). Prior to the in-situ annealing, the specimens were already partially recrystallized. Recrystallization boundaries with misorientation close to $40^{\circ}/\langle 111 \rangle$ were selected from preliminary EBSD scans over large areas using a large step size of $10~\mu$m and a detector binning factor of 4. Such boundaries were chosen because they were expected to migrate even at the relatively low annealing temperature of $50\,^{\circ}\mathrm{C}$, which is the maximum temperature attainable with this heating stage.

\subsection{Boundary migration patterns and in-plane strain analysis}

\begin{figure}[htbp]
  \centering
  \includegraphics[width=\textwidth]{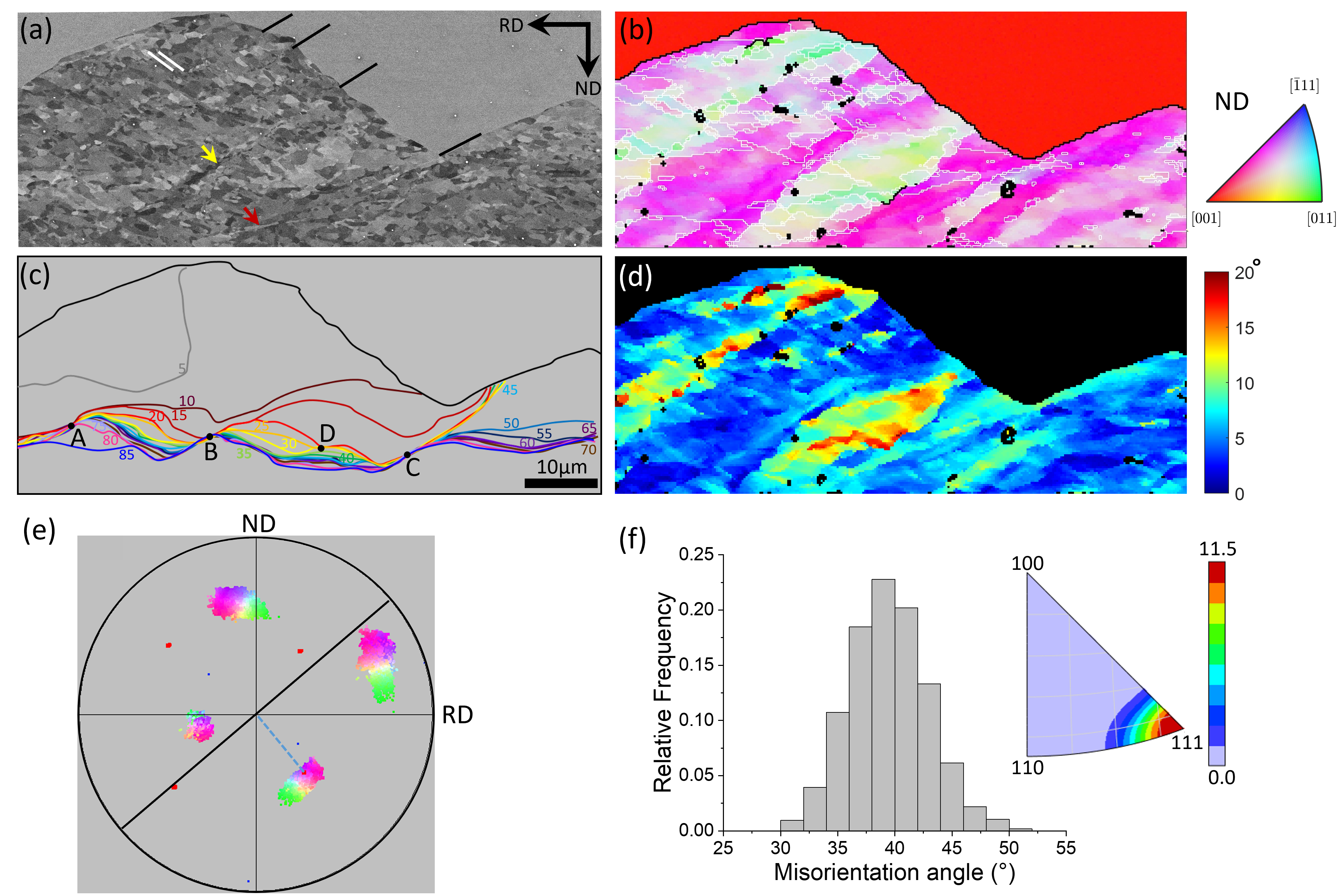}
  \caption{ECC image (a) and EBSD map (b) showing the microstructure of the initial partially recrystallized specimen containing a recrystallizing grain and deformed matrix. In (a), the yellow and red arrows mark examples of S-bands and lamellar bands, respectively, while the white and black lines mark examples of the narrowly spaced and irregularly spaced dislocation boundaries, respectively. In (b), the colors represent the crystallographic direction parallel to ND, while black pixels are non-indexed. (c) Traces of the position of the recrystallization boundary during the whole annealing duration, where for clarity only every fifth step (500~s intervals) is shown, as indicated by the numbers. (d) Deviation angle of the misorientation between the recrystallizing grain and each pixel in the deformed matrix to the ideal $\Sigma 7$ orientation relationship. (e) $\{111\}$ pole figure of the EBSD map in (b). The solid black line indicates the intersection line direction of the shared $\{111\}$ plane between the recrystallizing grain and the deformed matrix on the observation surface. (f) Distribution of misorientations between the recrystallizing grain and all the pixels in the deformed region. Panels (a)--(c) are modified from Ref.~\cite{ref47}.}
  \label{fig:fig1}
\end{figure}

For the first specimen, in-situ measurements were carried out using electron channelling contrast (ECC) imaging over an extended duration to quantify both the boundary migration patterns and possible in-plane shear-coupled motion by tracking the displacements of surface markers. The microstructures surrounding the selected boundaries were initially characterized by EBSD with a fine step size of $0.2~\mu$m and by ECC with a frame time of approximately 10~min to ensure high image quality. To monitor the migration of the recrystallization boundary in this specimen, hereafter referred to as GB1, the specimen was heated to $50\,^{\circ}\mathrm{C}$ in 2~min, and the migration of the recrystallizing boundary was subsequently recorded using ECC over 8600~s with a frame interval of 100~s, resulting in 87 boundary migration steps (see Fig. \ref{fig:fig1}). The short frame time was chosen to capture the detailed boundary migration, albeit at the expense of image quality in the deformed matrix.

\begin{figure}[htbp]
  \centering
  \includegraphics[width=\textwidth]{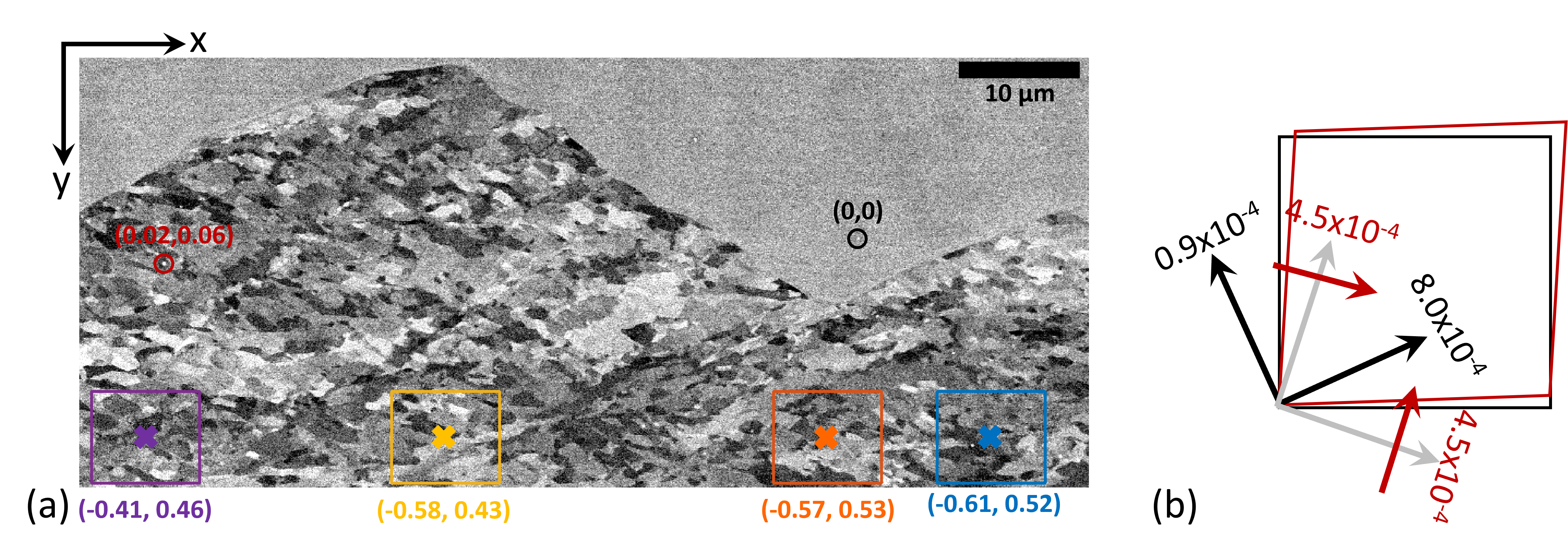}
  \caption{(a) Particles and deformed subregions selected for the in-plane strain analysis, together with their displacements relative to the top-right particle (marked by the black circle). The displacements are expressed in pixels, where 1 pixel corresponds to 80~nm. The crosses indicate the centers of the subregions, and the average displacements at these positions were used for strain calculation. (b) Sketch illustrating the in-plane strain tensor, determined from particle/subregion displacements that occurred during in situ annealing. The parallelogram illustrates the in-plane deformation of an initial square shape at a significantly exaggerated scale, while the associated black and grey vectors mark the directions of principal strains and maximum shear strains, respectively. Red arrows represent the maximum shear strains.}
  \label{fig:fig2}
\end{figure}

In-plane strains were analyzed by tracking the relative displacements, between the first and last annealing steps, of two small particles observed on the specimen surface and four subregions in the deformed matrix that remained identifiable throughout the annealing process (Fig.~\ref{fig:fig2}a). The particles were selected because they exhibited a good signal-to-noise ratio throughout the image sequence and were located as far as possible from the lower deformed subregions, thereby covering the region swept by GB1 during annealing. The relative displacements of these six reference points were fitted using an affine transformation, $\mathbf{A}$, from which the in-plane strain tensor, $\langle \varepsilon \rangle$, was determined as follows:
\begin{equation}
\langle \varepsilon \rangle = \frac{\mathbf{A} + \mathbf{A}^{\mathrm{T}}}{2} - \mathbf{I},
\end{equation}
where $\mathbf{I}$ is the identity matrix and the superscript $\mathrm{T}$ denotes the transpose operation. The resulting strain tensor is therefore representative of the average strain released during the entire in-situ annealing process over the area defined by the reference points.

For the subregions within the deformed matrix, displacement fields were determined using the global finite-element DIC approach~\cite{ref33}. To improve displacement accuracy to the subpixel level, a spacetime regularization scheme was applied to fit the displacement evolution of individual subregions over the whole image sequence~\cite{ref48,ref49}. For the temporal basis of the spacetime analysis, a B-spline function was used to ensure smooth evolution of displacement over time. The average displacements at the centers of the subregions, marked by crosses in Fig. \ref{fig:fig2}a, were then used to calculate the strains. In contrast, for the particles located in the recrystallizing grain, the local image contrast was insufficient for reliable DIC analysis. Their positions were therefore determined using intensity-weighted centroiding~\cite{ref50}, followed by a B-spline fitting to obtain subpixel positional accuracy. Further details of the analysis procedure and the associated error estimation are provided in Section~S1 of the Supplementary Material.

\subsection{Residual stress analysis}

For the second specimen, HR-EBSD was employed to quantify the local residual stresses around a selected recrystallization boundary, hereafter referred to as GB2. EBSPs were collected without detector binning before and after an annealing step at $30\,^{\circ}\mathrm{C}$ for 5~min. The patterns were acquired with an exposure time of 100~ms and stored on the local hard drive. Global (integrated) DIC~\cite{ref26,ref27} was employed to correlate the gradients of the experimental EBSPs with those of the simulated master pattern generated by EMsoft~\cite{ref51}; this approach is referred to as IDIC-G$\sigma$ in the following text. The projection center coordinates were linearly fitted to enhance precision. Elastic moduli $C_{11}=104$~GPa, $C_{12}=58$~GPa, and $C_{44}=28.8$~GPa~\cite{ref52} were used to calculate the in-plane stress components from the measured elastic strains, while the out-of-plane stress components were set to zero, corresponding to a plane-stress state at the sample surface. The EBSPs were indexed using IDIC-G$\sigma$ and processed using MTEX \cite{Bachmann2010MTEX}. The scanned area was separated into two parts, namely the recrystallizing grain and the deformed matrix. Further details of the IDIC-G$\sigma$ analysis can be found in Refs.~\cite{ref26,ref27}.

Following the prolonged EBSD scan required to obtain high-quality EBSPs during the first step, a thin carbon layer formed on the sample surface, degrading the pattern quality in the second step and hindering accurate IDIC-G$\sigma$ analysis (see also the representative patterns in Fig.~S4 in the Supplementary Material). Therefore, only the results from the starting condition are presented in the following.

\section{Results}

\subsection{Microstructure and boundary migration pattern for GB1}

The EBSD map and corresponding ECC image of the region surrounding GB1 are shown in Fig. \ref{fig:fig1}, together with schematic traces of the recrystallization boundary positions throughout the annealing process (Fig.~\ref{fig:fig1}c). For clarity, only every fifth step (corresponding to 500~s intervals) is displayed. While detailed quantification of local boundary migration velocities can be found in Ref.~\cite{ref47}, key microstructural features important for the understanding of local strain development are summarized below, together with a detailed analysis of boundary characteristics.

The recrystallizing grain exhibits an orientation of approximately $12^{\circ}$ from the ideal Cube orientation, $\{100\}\langle 001 \rangle$, while the average orientation of the deformed matrix is rotated about $20^{\circ}$ from the S, $\{123\}\langle 634 \rangle$, orientation (see Fig.~\ref{fig:fig2}a). The average misorientation between the recrystallizing grain and the deformed matrix is $38.8^{\circ}/\langle 0.61\,0.52\,0.60 \rangle$, deviating by $2.9^{\circ}$ from the ideal $\Sigma 7$ orientation relationship, though there exists a significant misorientation variation (approximately $15^{\circ}$) along the recrystallization boundary (see Figs.~\ref{fig:fig1}d and~\ref{fig:fig1}f).

Two distinct sets of dislocation boundaries are present in the deformed matrix. The first set consists of narrowly spaced dislocation boundaries (exemplified by the two white lines in Fig.~\ref{fig:fig1}a) oriented at angles of $-30^{\circ}$ to $-35^{\circ}$ to the sample rolling direction (RD), forming a microband (MB) structure. The second set is more irregularly spaced (examples are marked by black lines in Fig.~\ref{fig:fig1}a) and spans a broader range of angles from $15^{\circ}$ to $50^{\circ}$ to the RD. The second set has undergone a transition from microbands to S-bands and lamellar bands (e.g. as marked by the yellow and red arrows, respectively, in Fig.~\ref{fig:fig1}a) upon local shear~\cite{ref53}. Compared with the first set of MBs, the misorientation across these irregularly spaced bands is larger, as evidenced by the pronounced color changes across them in Fig.~\ref{fig:fig1}b. Among this set of bands, those appearing green-yellowish exhibit large deviation angles (up to $20^{\circ}$) from the ideal $\Sigma 7$ orientation relationship with respect to the recrystallizing grain (see Fig.~\ref{fig:fig1}d).

The migration pattern of the recrystallization boundary is heterogeneous. During the first 1000~s, boundary migration is influenced by the initial surface grain-boundary groove. In the subsequent, relatively steady stage, migration occurs predominantly along the irregularly spaced set of boundaries, and boundary segments parallel to this set of bands become pinning points (marked as A, B, C and D in Fig.~\ref{fig:fig1}c). As shown in Fig.~\ref{fig:fig1}e, the shared $\{111\}$ plane of the recrystallizing grain and the deformed matrix inclines about $50^{\circ}$ with respect to the observed surface, resulting in an intersecting trace line (its direction is indicated by the black line in Fig.~\ref{fig:fig1}e). Since the boundary segments moving along the irregularly spaced bands are roughly perpendicular to this trace line of the pure twist $\{111\}$ plane, they likely display a tilt-dominant characteristic. In contrast, the stationary segments at points A, B and C, which are oriented close to this intersecting trace line at these points, likely consist of a twist component. The observed faster migration of the tilt-dominant boundary segments than the twist ones agrees with classic observations~\cite{ref54,ref55}.

\subsection{In-plane strains for GB1}

Figure~\ref{fig:fig2} shows the relative displacements of the six reference positions with respect to the top-right particle from the first to the last annealing step, while the absolute displacements are shown in Fig.~S1 in the Supplementary Material. The relative displacement between the two particles at the top is rather small, on the order of 2--5~nm, whereas comparatively large displacements of around 40-50~nm, both horizontally to the left and vertically downward, are observed for the four selected subregions. These displacements correspond to an affine transformation matrix
\begin{equation}
\mathbf{A}=
\begin{bmatrix}
1.0006 & -0.0002 \\
-0.0004 & 1.0003
\end{bmatrix}.
\end{equation}
The normal strains along the RD and ND determined from $\mathbf{A}$ were found to be $\langle \varepsilon_{11} \rangle = 6.3 \pm 2.6 \times 10^{-4}$ and $\langle \varepsilon_{22} \rangle = 2.5 \pm 2.3 \times 10^{-4}$, respectively, accompanied by shear strains of $\langle \varepsilon_{12} \rangle = \langle \varepsilon_{21} \rangle = -3.0 \pm 2.2 \times 10^{-4}$. These strains predominantly develop while the recrystallizing grain replaces the deformed matrix and are measured relative to the starting microstructure, i.e. that immediately before the in-situ annealing. A principal strain analysis based on this in-plane strain tensor reveals large and small tensile strains roughly along and perpendicular to the irregularly spaced set of boundaries, respectively (see Fig.~\ref{fig:fig2}). The two maximum shear strain directions are approximately $61^{\circ}$ and $-29^{\circ}$ relative to the RD.

\subsection{Residual strain/stress for GB2}

\begin{figure}[htbp]
\centering
\includegraphics[width=\linewidth]{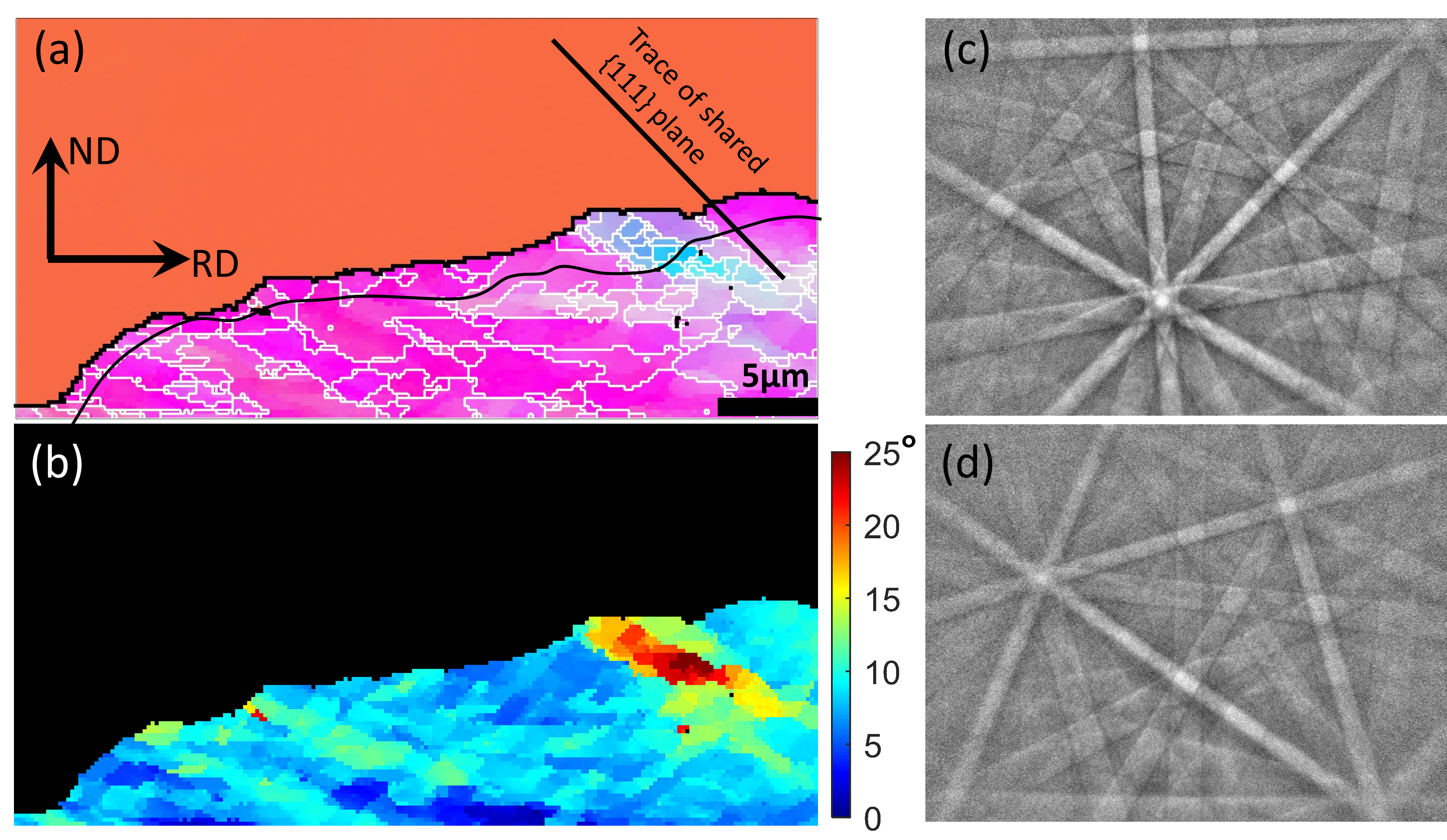}
\caption{EBSD results at GB2. (a) EBSD map colored according to the crystallographic orientation along ND. (b) Deviation angle of the misorientation between the recrystallizing grain and each pixel in the deformed matrix to the ideal $\Sigma 7$ orientation relationship. (c,d) Two exemplary EBSPs taken from the recrystallized and deformed grains, respectively.}
\label{fig:fig3}
\end{figure}

Figure~\ref{fig:fig3} shows the EBSD map across the second recrystallization boundary, i.e. GB2. The boundary trace after one annealing step is also indicated by the black line in Fig.~\ref{fig:fig3}a. Similar to the deformed matrix shown in Fig.~\ref{fig:fig1}, two sets of deformation bands are evident in the deformed region in front of GB2. Both the recrystallizing grain and deformed grain exhibit crystallographic orientations comparable to those observed for GB1, although the two deformed grains belong to different variants of the four possible equivalent S orientations. The misorientation across the boundary is close to the ideal $\Sigma 7$ orientation relationship, with the largest deviation of approximately $25^{\circ}$ inside the band that appears turquoise in Fig.~\ref{fig:fig3}a. The dominant set of deformation bands, which is parallel to the turquoise band and exhibits large misorientations across the bands, is aligned with the trace of the shared $\{111\}$ plane, which is inclined by approximately $45^{\circ}$ to RD, as indicated by the black line in Fig.~\ref{fig:fig3}a. The main boundary migration direction along this shared $\{111\}$ plane trace suggests that the faster-migrating segments possess a tilt component, consistent with the observations for GB1.

\begin{figure}[htbp]
\centering
\includegraphics[width=\linewidth]{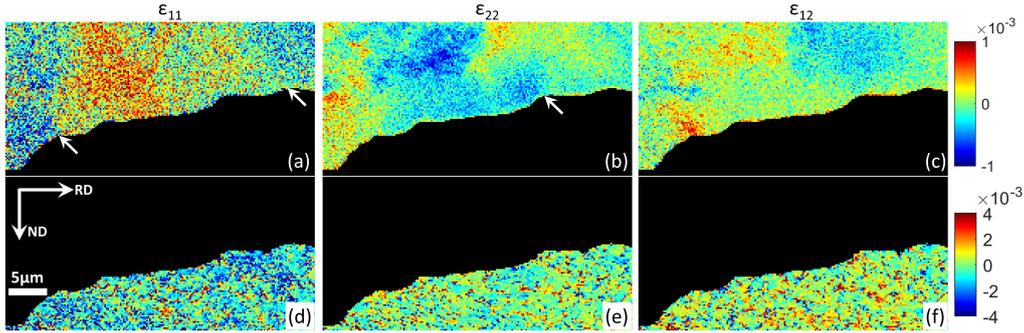}
\caption{In-plane elastic strain maps for (a--c) the recrystallizing grain and (d--f) the deformed matrix.}
  \label{fig:fig4}
\end{figure}

Representative EBSPs from the recrystallizing grain and the deformed grains are also displayed (Figs.~\ref{fig:fig3}c and~\ref{fig:fig3}d, respectively). As expected, the diffraction pattern quality from the recrystallizing grain is better than that from the deformed grain; however, both are of sufficiently high quality for the IDIC-G$\sigma$ analysis. The elastic strains determined by IDIC-G$\sigma$ for the recrystallizing grain and the neighboring deformed matrix are presented separately in Fig.~\ref{fig:fig4}. With the applied plane-stress state for the calculation, the resulting two out-of-plane shear strain components are much smaller than the in-plane shear component and are therefore not shown. The out-of-plane normal strain component is also omitted, as it is less critical for the present analysis and is of lower magnitude than the other two normal components.

In the recrystallizing grain, the strain components exhibit residual elastic strain variation on the order of $10^{-3}$ over a distance of approximately $10~\mu$m (Fig.~\ref{fig:fig4}a--c). The two in-plane normal elastic strains, $\varepsilon_{11}$ and $\varepsilon_{22}$, display a mixture of tensile and compressive regions and largely show opposite strain signs relative to each other. These strain value transitions are, to a certain extent, correlated with local boundary retrusions (as marked by the white arrows in Fig.~\ref{fig:fig4}a and~\ref{fig:fig4}b), indicating an influence of boundary shape on the strain distribution.

The elastic strains within the deformed grains are several times larger than those in the recrystallizing grain, with maximum values exceeding $\pm 5 \times 10^{-3}$. These values are typically associated with dislocation boundaries and are therefore likely not very accurate, as EBSPs acquired in these regions often exhibit overlapping patterns, which deteriorates the IDIC-G$\sigma$ analysis. These strain values are therefore excluded in the following analysis. In contrast to the recrystallizing grain, the in-plane normal strain $\varepsilon_{11}$ is predominantly compressive, with an average value of $-6.5 \times 10^{-4}$, whereas the normal strain $\varepsilon_{22}$ exhibits both compressive and tensile regions, resulting in an average value close to zero. The influence of the two sets of deformation bands on the elastic strain distribution can also be noticed, albeit not very pronounced.

\begin{figure}[!htbp]
\centering
\includegraphics[width=0.7\linewidth]{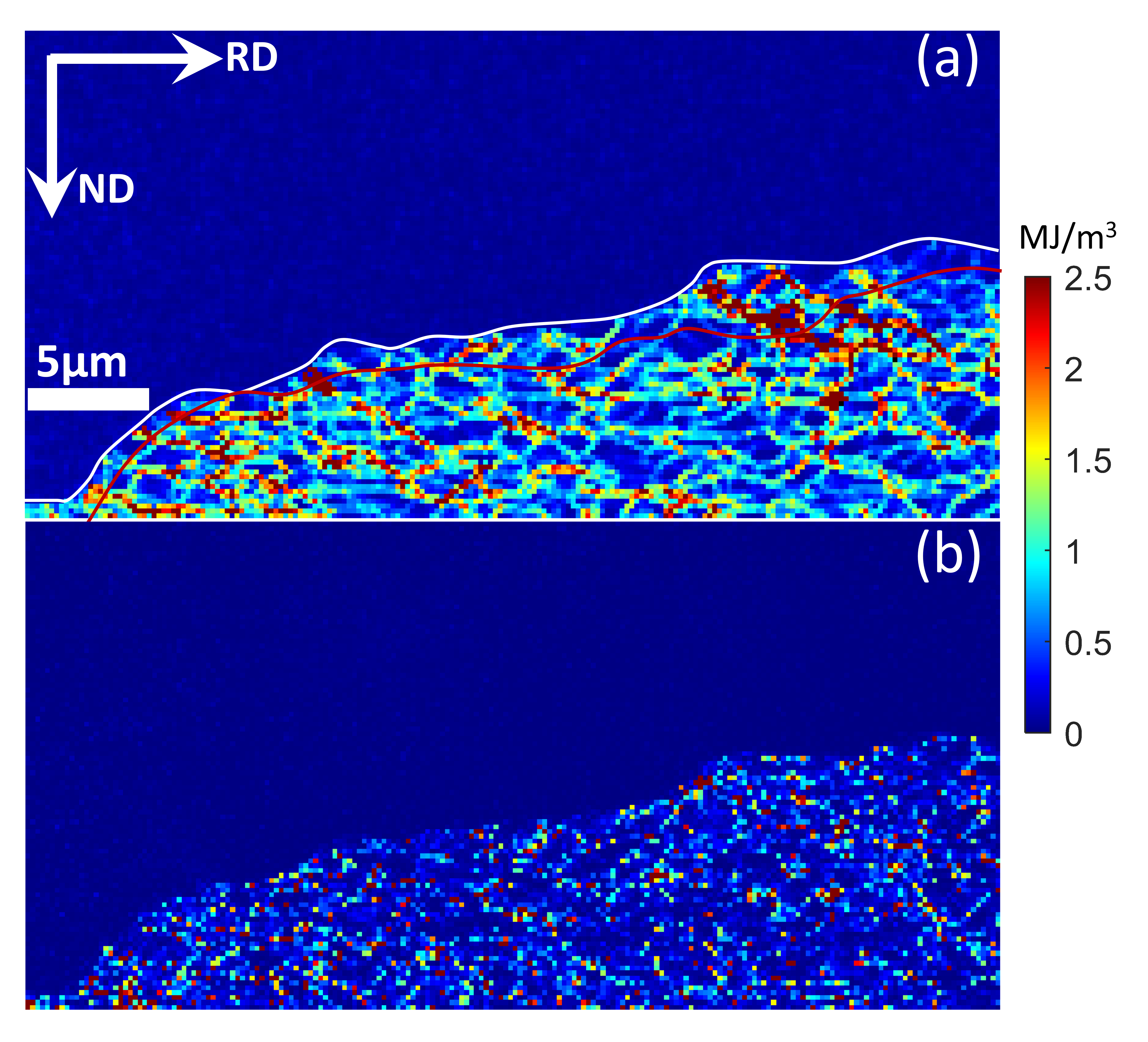}
\caption{Stored energy distribution: (a) plastic strain contribution determined from KAM values; (b) elastic strain energy calculated from the elastic strain–stress products. The boundary traces are overlaid in (a) to highlight the migration distance. Black pixels in (b) correspond to strain magnitudes exceeding 0.005.}
\label{fig:fig5}
\end{figure}

The stored energy distribution determined from the kernel average misorientation (KAM) values and from the elastic strains/stresses, representing plastic and elastic strain energies, respectively, is shown in Fig.~\ref{fig:fig5}. KAM values, $\theta_{\mathrm{KAM}}$, were obtained from the HR-EBSD analysis, and no lower cut-off misorientation angle was used for the calculation, while the method described in Ref.~\cite{ref56} was applied for calculating the plastic strain energy, $E_{\mathrm{p}}$, according to

\begin{equation}
E_{\mathrm{p}} = \frac{3Gb\theta_{\mathrm{KAM}}}{2\Delta},
\end{equation}

where $G$ is the shear modulus, $b$ is the Burgers vector, and $\Delta$ is the step size. For the elastic strain energy, $E_{\mathrm{e}}$, a sum of the stress--strain product over all components was used~\cite{ref57}, i.e.

\begin{equation}
E_{\mathrm{e}} = \frac{1}{2}\sigma_{ij}\varepsilon_{ij}.
\end{equation}

The elastic strain energy is lower than the plastic strain energy estimated from the KAM values. The average elastic strain energy remains approximately one-half of the corresponding plastic strain energy, 0.31 versus $0.67~\mathrm{MJ\,m^{-3}}$. Overall, no significant variation in stored energy is seen along the boundary, except in the vicinity of the band shown in turquoise in Fig.~\ref{fig:fig3}a, where locally elevated plastic stored energy values of up to $7~\mathrm{MJ\,m^{-3}}$ are detected.

\section{Discussion}

\subsection{Comparison of in-plane strains and HR-EBSD residual strains}

\begin{figure}[!htbp]
  \centering
  \includegraphics[width=0.7\linewidth]{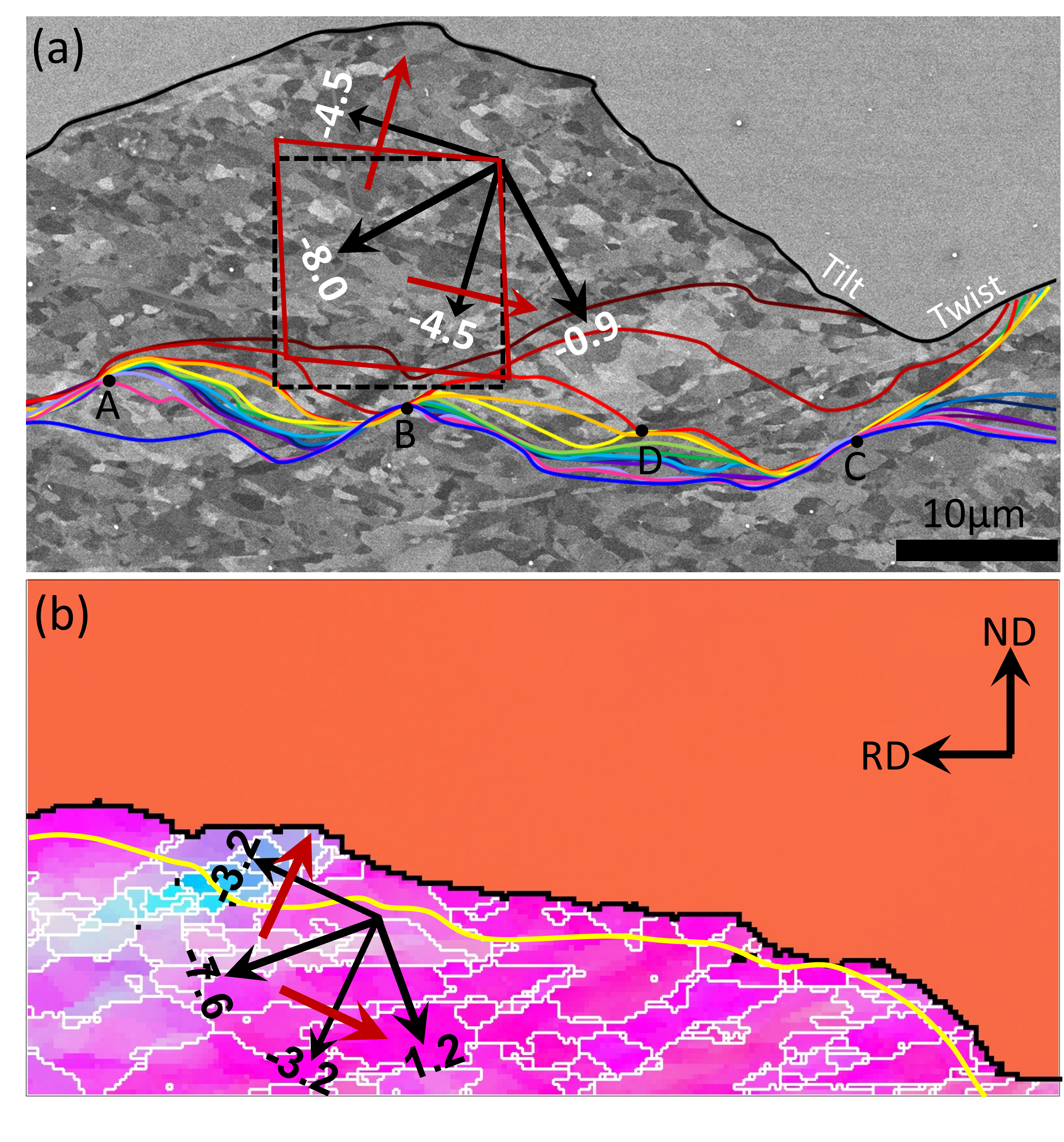}
  \caption{Principal strain/stress analysis and their relationship to local boundary migration patterns. (a) Average principal strain tensor for the deformed matrix in front of GB1, superimposed on part of the deformed matrix and recrystallization boundary traces. The red parallelogram illustrates the residual in-plane deformation matrix within the deformed grain, referenced to the dashed black square at a significantly exaggerated scale. (b) Average principal in-plane strain directions for the deformed matrix and their relative orientations with respect to the geometrical alignment of the dislocation boundaries, overlaid on the EBSD map. The arrows with thick and thin black lines indicate the principal strain and maximum shear strain directions, respectively, while the red arrows denote maximum shear strains. Numbers denote strain values, with the $\times 10^{-4}$ factor omitted. Panel (b) is flipped about ND for comparison with (a).}
  \label{fig:fig6}
\end{figure}

The in-plane strain tensor obtained using the combined particle-tracking and global DIC methods for GB1 represents the relative strain that developed as the recrystallizing grain replaced the deformed matrix. Assuming an average zero-strain state for the recrystallizing grain at the end of the in-situ annealing, the measured affine transformation matrix corresponds to a release of the residual strains in the deformed grain (as illustrated by the red parallelogram in Fig.~\ref{fig:fig6}a). Accordingly, the principal strain with maximum magnitude in the deformed grain is estimated to be $-(8.0 \pm 2.6) \times 10^{-4}$, aligned at about $-17^{\circ}$ relative to RD, along roughly the dominant boundary migration direction (see Fig.~\ref{fig:fig6}a).

For GB2, if the average strain values determined by IDIC-G$\sigma$ for each in-plane strain component in the deformed matrix are used to calculate the average principal strain directions, the resulting strains are $-7.6 \times 10^{-4}$ along a direction approximately $-20^{\circ}$ from RD, $1.2 \times 10^{-4}$ perpendicular to this direction, and a shear strain of $-3.2 \times 10^{-4}$ oriented at approximately $25^{\circ}$ and $115^{\circ}$ relative to RD. This average simulates the mean strain state experienced as the boundary sweeps through the corresponding area. This result is illustrated in Fig.~\ref{fig:fig6}b, where both the microstructure and the strain tensor are flipped about ND to enable direct comparison with GB1 in Fig.~\ref{fig:fig6}a.

The magnitude and direction of the two average strain tensors shown in Fig.~\ref{fig:fig6}a and~\ref{fig:fig6}b are in very close agreement, which validates the mutual consistency and accuracy of the two methods used. More importantly, the combined results clearly demonstrate the involvement of local residual stresses during recrystallization boundary migration. Although the out-of-plane normal stress components are zero at the free surface in HR-EBSD, the analyses unambiguously reveal the distinct local variations in the residual stress field. These results provide a sound basis for discussing the development of residual stresses within recrystallizing grains and their influence on local boundary migration behavior, as well as on shear-coupled migration mechanisms for recrystallization boundaries.

\subsection{Development of residual stresses}

The HR-EBSD results clearly reveal the presence of local residual stresses within the recrystallizing grains, consistent with previous observations~\cite{ref2,ref3,ref4,ref5,ref6}. However, both the residual strain levels and their spatial variations are higher than those previously reported for Al alloys~\cite{ref6,ref58} and other alloys~\cite{ref2,ref3,ref4,ref5}. The elevated residual stresses in the recrystallizing grains are attributed to the even higher residual stresses in the surrounding deformed matrix (see Fig.~\ref{fig:fig4}). Such residual stresses in the deformed grains typically arise from imperfect dislocation patterning associated with the generation and storage of dislocations during plastic deformation~\cite{ref59,ref60}. In the present sample, which was cold rolled at liquid-nitrogen temperature, this effect is further amplified compared with deformation at room temperature, leading to larger medium- and long-range residual stresses within the deformed matrix.

The difference between the average values of the two in-plane normal strain components in the deformed grain (e.g. $\langle \varepsilon_{11} \rangle = -6.5 \times 10^{-4}$ and $\langle \varepsilon_{22} \rangle = 0.1 \times 10^{-4}$ for GB2) suggests that the residual stresses are influenced by the macroscopic rolling process applied to the sample. This is in good agreement with observations in cold rolled iron and gum metal \cite{ref3,ref5}. 
Moreover, as shown in Figs.~\ref{fig:fig6}a and~\ref{fig:fig6}b, the average maximum principal (compressive) strain directions are closely aligned with the dominant set of deformation bands, indicating that the strain field is strongly influenced by the local microstructure, particularly by the geometrical arrangement of these bands.

The average strains in the recrystallizing grain for GB2 are close to zero: $\langle \varepsilon_{11} \rangle = 0.6 \times 10^{-4}$ and $\langle \varepsilon_{22} \rangle = -1.0 \times 10^{-4}$. The lower strain magnitudes in the recrystallizing grain compared with those in the deformed matrix are partly attributed to the relatively low defect density of recrystallizing grains and, hence, their higher physical density compared with the deformed matrix~\cite{ref63,ref64}. When recrystallizing grains replace the deformed matrix, this density difference imposes, on average, a tensile hydrostatic strain within the recrystallizing grains~\cite{ref5}, thereby releasing the hydrostatic compressive strain previously present in the replaced deformed matrix. In addition, the intragranular strain variations observed in the recrystallizing grain (Fig.~\ref{fig:fig4}a--c) indicate that other factors, including interactions with surrounding deformed grains and the local grain-boundary geometry, also influence the development of local residual strains. In future work, the interaction between the residual strains in recrystallizing grains and those in the surrounding matrix should be quantified.

\subsection{Impact of residual stresses on boundary migration}

\subsubsection{Shear-coupled migration}

The presence of local residual stresses can result in shear stresses across the boundary. However, the shear component of the in-plane affine transformation matrix for GB1 indicates a maximum shear-coupling factor of $3.0 \times 10^{-4}$, which is very small. During the in-situ annealing experiment, only an in-plane vertical displacement of approximately $5$~nm was observed between the two selected particles, which were approximately $50~\mu$m apart. This small displacement is therefore more likely to result from the relaxation of elastic strain stored in the deformed matrix than from shear-coupled boundary migration.

\begin{figure}[htbp]
  \centering
  \includegraphics[width=0.65\linewidth]{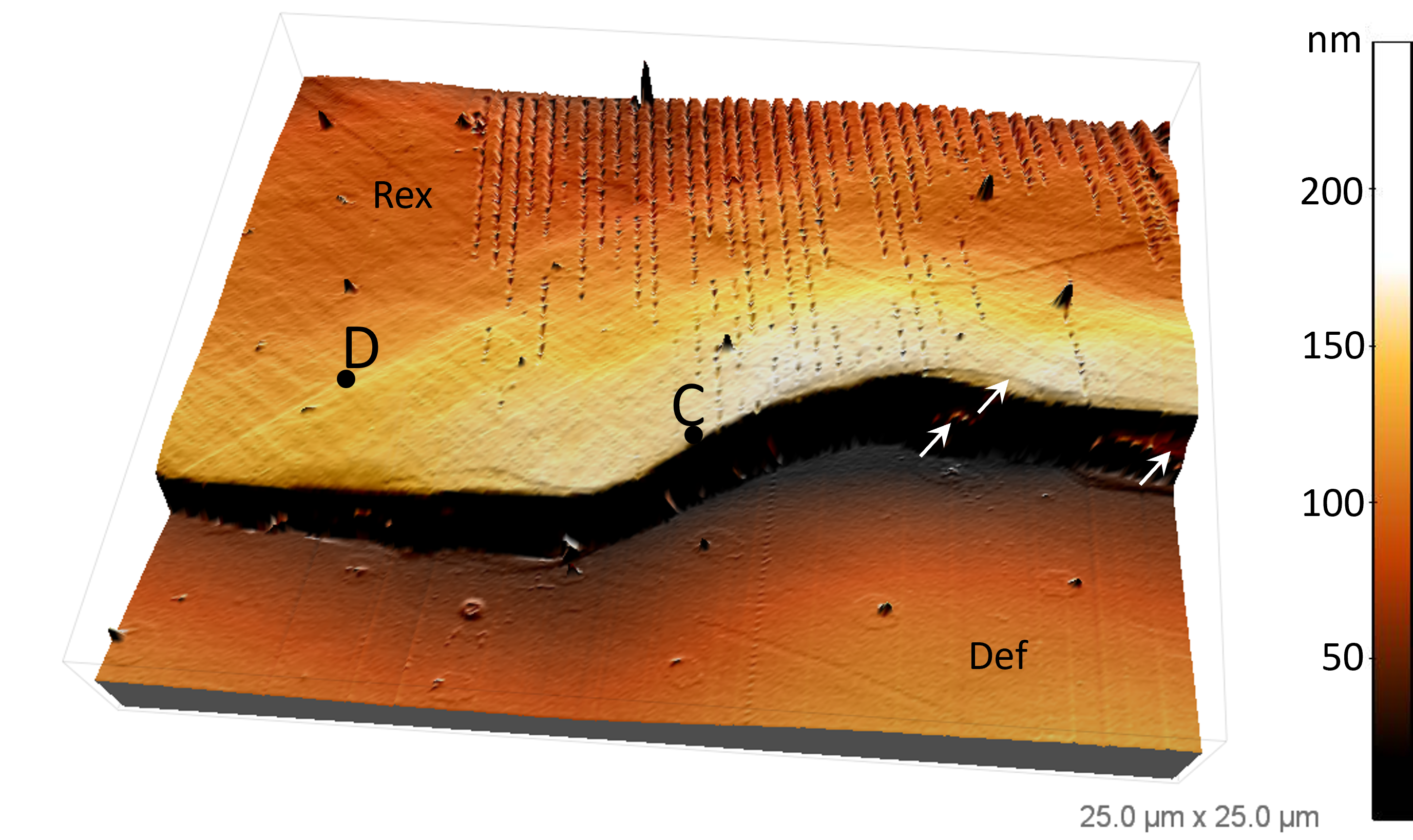}
  \caption{Surface morphology measured by AFM in a region across GB1, ten months after the in-situ measurements. White arrows indicate some small staircase steps. Letters C and D identify the same pinning locations as those marked in Fig.~1c. This figure is modified from Ref.~\cite{ref65}.}
  \label{fig:fig7}
\end{figure}

Since the observations were made on the free sample surface, one may speculate whether shear-coupled migration could occur out of plane. However, atomic force microscopy (AFM) analysis of the surface morphology does not support this interpretation. Within the region swept by the recrystallizing grain during the in-situ annealing experiment, the out-of-plane surface displacement associated with short-time stagnation of the boundary segments was only about $2$~nm as detailed in Ref.~\cite{ref65}. In contrast, a distinct staircase feature with a height difference of $150$--$200$~nm is observed across the recrystallization boundary at its final position (Fig.~\ref{fig:fig7}). Closer inspection shows the presence of smaller staircase steps within the larger feature, as marked by the white arrows in Fig.~\ref{fig:fig7}. These steps coincide with boundary traces where the boundary segments remained stationary for certain periods (see Fig.~\ref{fig:fig1}c). Since the AFM measurements were carried out ten months after the in-situ experiment~\cite{ref65}, the large height difference most likely developed during that interval while the boundary remained \textit{stagnant} and is therefore not associated with shear-coupled \textit{migration}.

In bulk samples, the growth of recrystallizing grains is constrained in all three dimensions, analogous to the in-plane confinement observed here. Under such conditions, shear-coupled migration is unlikely to operate during recrystallization boundary motion, and boundary migration is therefore more consistent with predominantly normal, diffusion-mediated boundary migration.

\subsubsection{Influence of residual stresses on local migration patterns}

Based on the stored energies and in-situ data, average grain boundary mobilities can be quantified; see, for example, Section~S3 of the Supplementary Material. However, fully understanding the local boundary migration patterns remains challenging. For GB2, the relatively large migration distance observed around the turquoise band is likely attributed to the high stored plastic energy in that region. This is particularly significant because the misorientations of the boundary segments in that area show larger deviations from the ideal $\Sigma 7$ boundary misorientation than the neighboring segments, implying a lower intrinsic boundary mobility~\cite{ref66}, whereas the elastic strain energy is distributed relatively uniformly across the deformed matrix.

In contrast, the pinned segment on the left side of this boundary (Fig. \ref{fig:fig3}a) is more difficult to explain. The stored energy in the deformed region ahead of this segment does not appear to be systematically lower than that of other segments along the boundary. The deviation of the boundary misorientation from the ideal $\Sigma 7$ relationship is even smaller than that of its neighboring moving segments, which should, in principle, correspond to a higher mobility. Previous studies have shown that the boundary misorientation rotation axis can have a stronger influence on mobility than the misorientation angle~\cite{ref67,ref68}, and that for $\Sigma 7$-type boundaries, mobility decreases as the rotation axis deviates from $\langle 111 \rangle$~\cite{ref66}. However, this is also not the case for the pinned segment: the misorientation rotation axis is deviated by a similar amount compared to the neighboring migrating segments (see Fig.~S6 in the Supplementary Material). A similar situation is observed for GB1 at positions A, B, and C in Fig.~\ref{fig:fig1}.

For both boundaries, the presence of dominant deformation bands certainly leads to bands of high local stored energies along their directions, which can promote preferential boundary migration along these bands, as demonstrated by phase-field simulations. This is because this migration pathway releases stored energy most effectively~\cite{ref69}. Nonetheless, when two sets of asymmetrical deformation bands are present, phase-field simulations do not predict boundary segments to remain stationary for extended periods, as observed at points A, B and C for GB1 in Fig.~\ref{fig:fig1}. These mixed behaviors between the moving and non-moving segments also indicate that the previously developed iterative mobility-matching method~\cite{ref68}, based solely on misorientation (angle/axis pair), needs to be extended to account for additional missing factors.

The presence of a residual strain/stress tensor could certainly be one of the missing factors. When considering the elastic energy from the strain tensor, it does not have any orientation dependence; that is, the elastic energy can be released equally efficiently regardless of the direction from which a given volume is consumed. However, the elastic strain itself is a tensor quantity and is therefore anisotropic with respect to different sample directions. It is thus important to examine whether this anisotropy can be linked to the local grain boundary migration patterns, especially the migrating and stagnant boundary segments are closely related to the principal strain directions in the deformed matrix (Fig.~\ref{fig:fig6}). In particular, the compressive principal strain appears to favor boundary migration. A similar behavior was observed in another pure Al sample~\cite{ref6}, where the recrystallizing grain exhibited a compressive strain along the growth direction, parallel to the dominant dislocation-boundary direction.

\begin{figure}[htbp]
  \centering
  \includegraphics[width=0.7\linewidth]{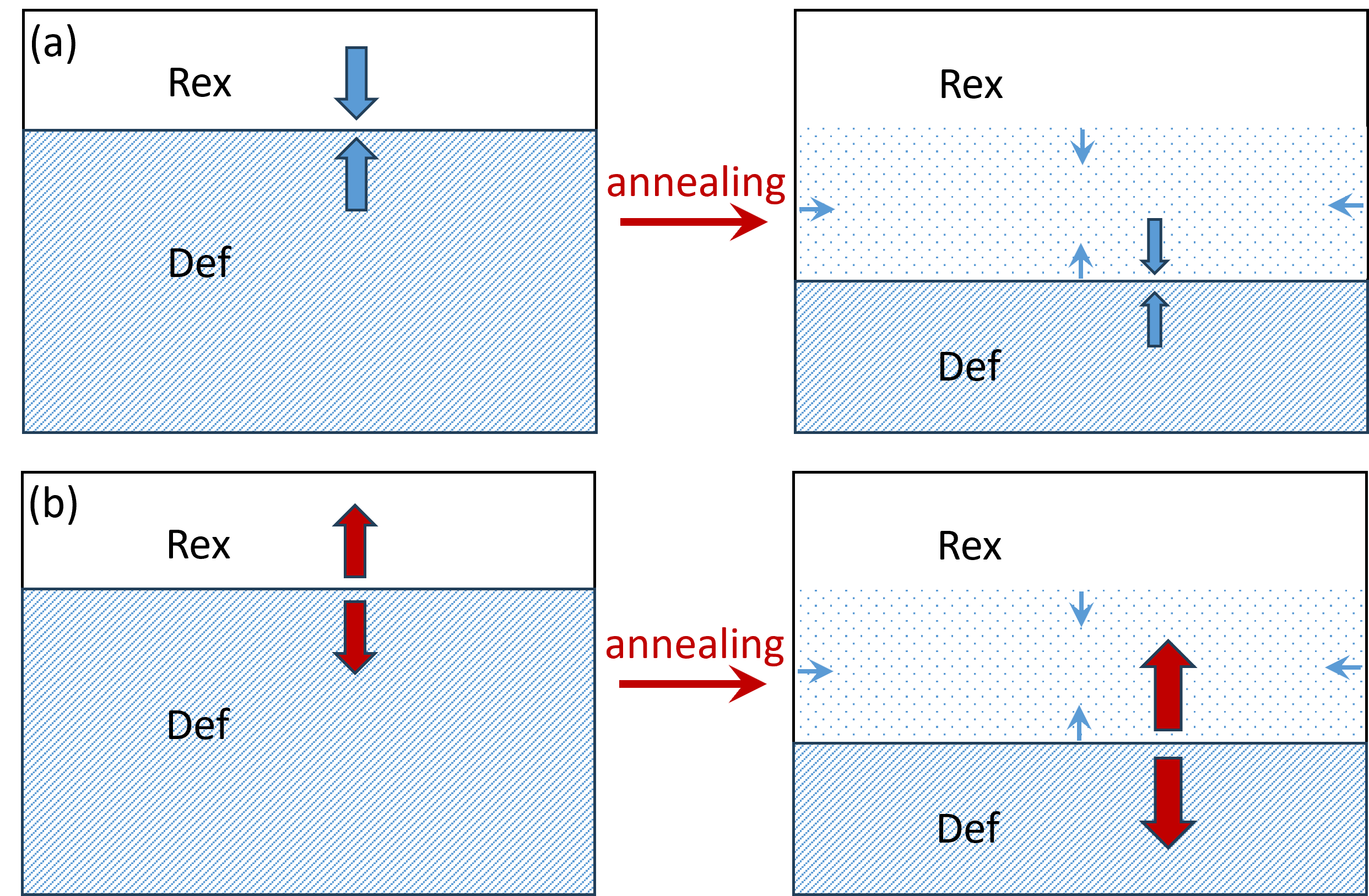}
  \caption{Schematic illustration of the interaction between residual strain and volume shrinkage generated as a recrystallizing grain consumes the surrounding deformed matrix, for the same recrystallization boundary under predominantly (a) compressive and (b) tensile residual strain conditions along the migration direction. The dotted region represents the newly recrystallized region formed during annealing. The large blue and red arrows represent tensile and compressive strains, respectively, with their sizes roughly proportional to the strain magnitude. The small arrows indicate the volume shrinkage.}
  \label{fig:fig8}
\end{figure}

When a recrystallizing grain consumes part of the deformed microstructure, a small overall volume shrinkage occurs. If the deformed matrix is initially under predominantly compressive strain along the migration direction, this shrinkage naturally reduces the residual strains on both sides of the boundary (see Fig.~\ref{fig:fig8}a). By contrast, if the deformed matrix is initially in tension, the volume shrinkage further increases the tensile strain on both sides (Fig.~\ref{fig:fig8}b), making the process less favorable in terms of reducing the overall stored energy. Thus, in addition to the energy stored within the consumed volume, the resulting change in the total elastic energy is also important. 

Moreover, previous studies on thin films have shown that stress fields can influence vacancy-mediated diffusion processes~\cite{ref70,ref71,ref72,ref73}, and similar behavior has been reported in Ni--Al alloys during elastic strain-assisted aging~\cite{ref74}. It has been proposed that stress fields distort the crystal lattice, thereby altering the distance over which atoms jump to the nearest vacancy. Specifically, a compressive strain reduces this distance along the strain direction, lowering the activation energy for atomic jumps and thus accelerating diffusion compared with unstrained conditions. Conversely, a tensile strain increases this distance, potentially slowing diffusion~\cite{ref74}.

Considering that grain boundary migration is a diffusion-controlled process, where atoms transfer from one grain to another across the boundary, it is reasonable to infer that strain/stress fields affect the boundary mobility in a similar manner. In the present case, the residual compressive strains in the recrystallizing grains along the dominant set of bands for both GBs are expected to enhance the boundary mobility. For the pinned segments, the residual strain component perpendicular to the boundary is comparatively small, whether tensile or compressive (Figs.~\ref{fig:fig6}a and~\ref{fig:fig6}b), thereby weakening this stress-enhancement effect.

Furthermore, in addition to local residual stresses, the 3D boundary-plane normal and the boundary tilt--twist character may play a role. For the present recrystallization boundaries, the migrating segments exhibit a tilt-dominant character, with typically higher mobility than twist boundaries of the same misorientation. A mobility difference of approximately one order of magnitude ($\sim 10$) is generally expected between tilt and twist $\Sigma 7$ boundaries~\cite{ref55,ref66}, which may contribute to the slow motion across the twist-dominant segments at points A, B, and C. However, the fastest-migrating segments also display larger deviation angles from the ideal $\Sigma 7$ orientation relationship (e.g. between B and C in Fig.~\ref{fig:fig1}c), although such deviations are generally associated with mobilities that are about an order of magnitude lower~\cite{ref66}. In addition, stress-dependent mobility has also been reported for ideal symmetrical tilt $\Sigma 7$ boundaries~\cite{ref46}, i.e.\ in the absence of a twist component. Similarly, twist-dominant boundary segments were observed to migrate in a pure Al sample~\cite{ref6}. Altogether, these observations suggest that the distinction between tilt and twist character may not be critical in this case. The combined influence of these crystallographic factors and the principal stresses should be further quantified across more boundaries.

\section{Conclusions}

Combined \textit{in situ} microstructure-based DIC and HR-EBSD analyses demonstrate that local residual stresses affect recrystallization boundary migration significantly in cryogenically rolled pure aluminum. The following conclusions can be drawn:

\begin{enumerate}
  \item Residual strains on the order of $10^{-3}$ are detected within recrystallizing grains, whereas strains in the adjacent deformed matrix are several times higher. The stresses observed within the recrystallizing grains are a result of the redistribution of medium- to long-range residual stresses originally stored in the deformed matrix, which are associated with the geometrical arrangement of the dislocation boundaries and originate from imperfect dislocation screening during cryogenic deformation.

  \item The spatial variation of these residual stresses within the recrystallizing grain suggests that their formation is affected by interactions with neighboring deformed grains, local grain-boundary geometry, and defect-density differences relative to the deformed matrix. In contrast, no pronounced influence of the deformation-band structure on the spatial distribution of local residual stresses within the deformed grain was observed.

  \item No evidence of shear-coupled boundary motion has been detected during boundary migration, despite the presence of significant local residual stresses. This suggests that migration of recrystallization boundaries is primarily normal to the boundary plane and controlled by diffusion, rather than by shear-driven mechanisms.

  \item The dominant deformation bands govern the principal residual strain field in the deformed matrix, which in turn strongly influences the local migration pattern. In particular, local boundary migration is promoted by the compressive principal strain, preferentially along these bands.
\end{enumerate}

Overall, the results provide direct experimental evidence that stress anisotropy and inherited residual stresses can affect recrystallization boundary migration, highlighting the need for future in-situ studies to quantify the coupling between stress evolution and migration at elevated temperatures. This work also highlights the crucial role of multimodal characterization in comprehensively understanding the complex recrystallization process.

\section*{Data availability}

Processed data will be available on request.

\section*{Acknowledgements}

The author thanks Prof.~Dorte Juul Jensen and Prof.~Andy Godfrey for their valuable comments and discussions on this work.

\section*{Funding}

YZ acknowledges support from the VILLUM FONDEN through the research grant MicroAM-VIL54495. SQ and GW acknowledge support from the National Natural Science Foundation of China (Grant Nos.~52273229 and 52471031).

\bibliographystyle{elsarticle-num}
\bibliography{cas-refs}

\end{document}